\documentclass[12pt]{iopart}

\usepackage{graphicx}  
\usepackage{hyperref}

\usepackage{epsfig}  
\usepackage{bm}  
\usepackage{color}  
\usepackage{float}  
\usepackage{dcolumn}   
\usepackage{multirow}   
\usepackage{longtable}  
\def\be{\begin{equation}}  
\def\ee{\end{equation}}

\begin{document}

\title{Challenges in Nuclear Structure Theory}  
  
\author{W. Nazarewicz}   

\address{Department of Physics and Astronomy and FRIB Laboratory, 
Michigan State University, East Lansing, Michigan  48824, USA}
\address{Institute of Theoretical Physics, Faculty of Physics,
University of Warsaw, 02-093 Warsaw, Poland}

\begin{abstract}  
The goal of nuclear structure theory is to build a comprehensive microscopic
framework in which properties of nuclei and extended nuclear matter, and nuclear reactions and decays can all be consistently described. 
Due to  novel theoretical concepts, breakthroughs in the experimentation with rare isotopes,
increased exchange of ideas across different research areas, and the progress in computer
technologies and numerical algorithms, nuclear theorists have been
quite successful in solving various bits and pieces of the nuclear
many-body puzzle and the prospects are exciting. This article contains  a brief, personal perspective on the status of the field.
\end{abstract}  
  
\maketitle  

\section{Introduction: the territory}

\begin{figure}[htb]
\center
\includegraphics[width=0.8\textwidth]{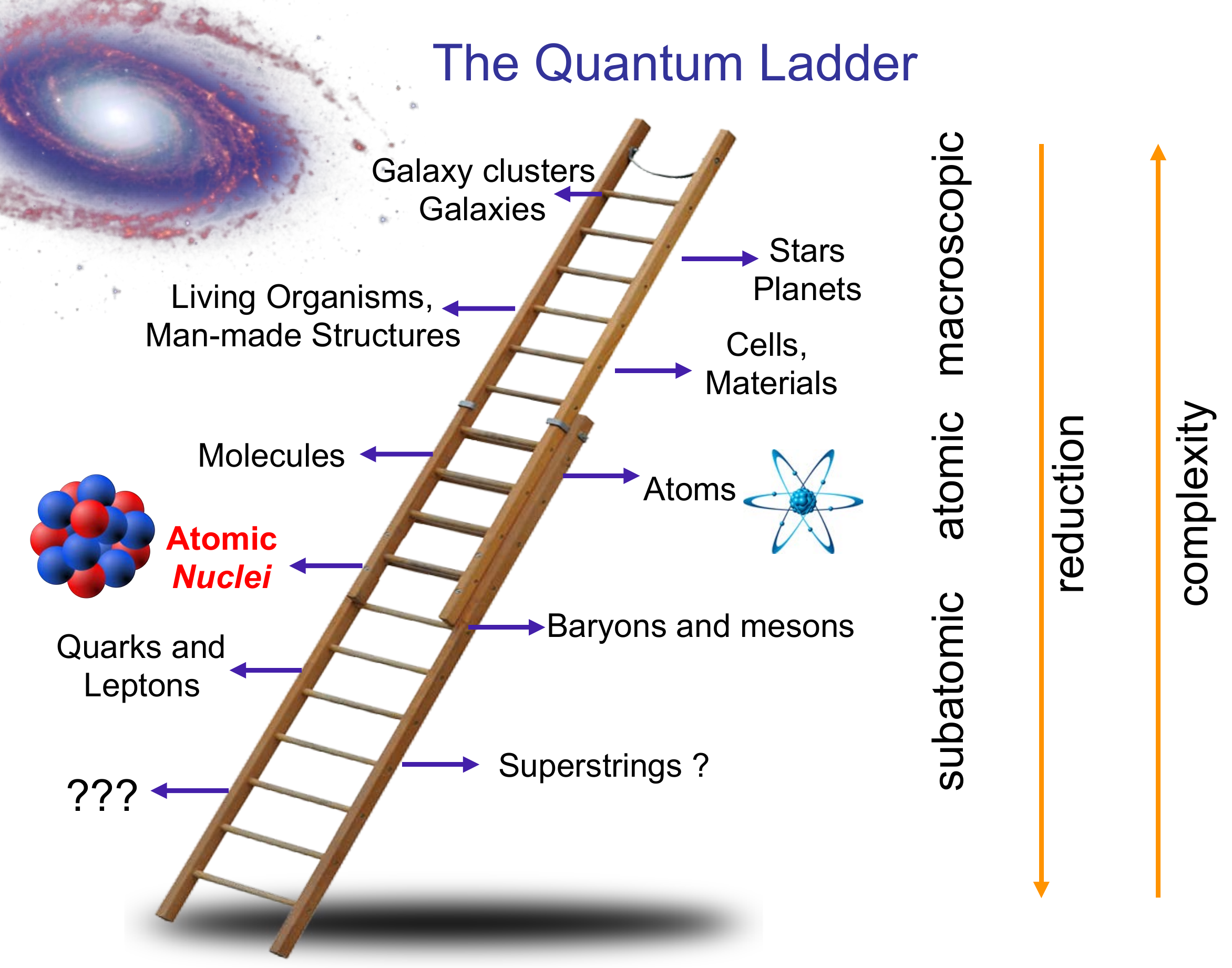}
	\caption{\label{ladder}
Quantum ladder:  physical systems at various scales, from microscopic to macroscopic.
As the atomic nucleus occupies the threshold  between the fundamental and the emergent,  nuclear structure physics is in a unique position to address overarching  science  questions from different perspectives. 
}
\end{figure}

The atomic nucleus is  placed  at the center of the quantum ladder: it provides a connection between the smallest and the largest, see Fig.~\ref{ladder}. Being a congregation of  neutrons and protons, it  emerges from complex interactions between quarks and gluons on a scale of femtometers. But its properties also determine the behavior of giant stars on a gigameter scale. Indeed, nuclear structure  encompasses phenomena over an incredibly wide range of energies and distances. Atomic nuclei are laboratories of fundamental laws of nature, and thus linked to particle physics. At the same time, nuclei exhibit behaviors that are emergent in nature and present in other complex systems studied by condensed matter physicists and  quantum chemists. 
The general challenge for the interdisciplinary field of nuclear structure -- the nuclear many-body problem -- is to understand the principles of building up nuclear complexity out of fundamental  degrees of freedom, which, when inspected at higher resolution, have  a complicated structure of their own.

The strategic location of the atomic nucleus on the quantum ladder is reflected in the overarching questions that drive the field \cite{Decadal2012,NSACLRP2015}: 
\begin{itemize}
\item
Where do nuclei and elements come from?
\item
How are nuclei made and organized?
\item
How can nuclei be exploited to reveal the fundamental symmetries of nature?
\item
What are practical and scientific uses of nuclei?
\end{itemize}
Complete  answers to these questions require much deeper understanding of atomic nuclei  than currently available. 
Both low-energy experimentation and nuclear structure theory answer these questions in a synergic manner.

\section{The Roadmap: Who could have predicted this 20 years ago?}

\begin{figure}[htb]
\center
\includegraphics[width=0.5\textwidth]{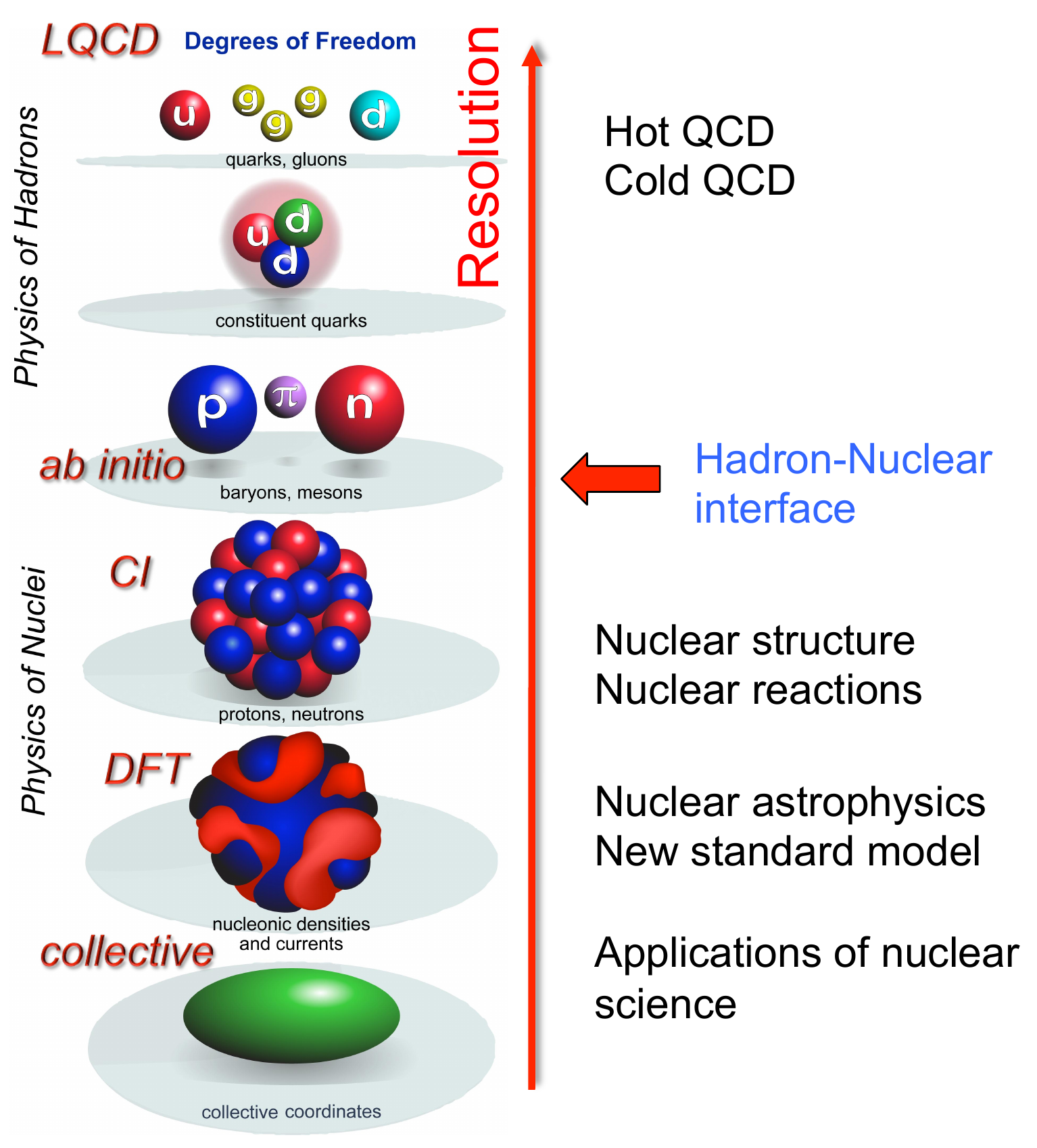}
	\caption{\label{DoF}
The basic elements (degrees of freedom) that atomic nuclei are made of depend on the energy of the experimental probe and the distance scale. The building blocks of quantum chromodynamics are quarks and gluons, which are lurking inside mesons and baryons. In low-energy nuclear physics experiments, nuclei can be well described in terms of individual protons and neutrons, their densities and currents, or -- for certain nuclear excitations -- collective coordinates describing rotations and vibrations of the nucleus as a whole. Major theoretical approaches to the nuclear many-body problem (Lattice QCD, {\it ab-initio} models,  configuration interaction techniques, nuclear Density Functional Theory, and collective model) are marked at different resolution scales
(Adopted from Ref.~\cite{(Ber07)}.)
}
\end{figure}

These are exciting times for theoretical nuclear structure
research \cite{Decadal2012,NSACLRP2015,NSACLRP2007}.  Figure~\ref{DoF}  shows the degrees of freedom of nuclear structure research in the context of  the theoretical roadmap of the nuclear many-body problem \cite{riatheory,(Ber07)}. Important insights come from the realization that the choice of nuclear building blocks always depends on  the resolution of the theoretical microscope (i.e., a specific model applied to a  specific physics question). In the following, the hierarchy of nuclear structure approaches is briefly discussed, in the direction of decreasing accuracy and increasing mass.
The references and highlights cited  are representative but not inclusive, as this perspective is not supposed to be a comprehensive review. For an excellent list of current achievements in nuclear structure theory, the reader is referred to recent reports \cite{Decadal2012,NSACLRP2015,LETM2014}.

Quantum chromodynamics (QCD), the underlying theory of strong interactions, governs the dynamics and properties of quarks and gluons that form baryons and mesons; hence, it is also responsible for the complex inter-nucleon forces that bind nuclei.
In this area, significant progress is being made by computing properties of 
the lightest nuclei and nuclear interactions within Lattice-QCD (LQCD) \cite{(QCD),(Doi13),(Bea15),(Cha15)}. 

The  nuclear problem with protons and neutrons is an effective approximation to QCD.
Here, effective field
theory (EFT) has enabled us to construct high-quality two-body and three-body inter-nucleon
interactions consistent with the chiral symmetry 
\cite{(Bea02),(Epe09),(Mac11),(Epe15)}. A low-momentum
inter-nucleon interaction has been derived using the similarity renormalization
group technique \cite{(Bog07),(Heb12)}. 
The low-energy coupling constants that appear in the chiral EFT expansion,  which incorporate unresolved short-distance physics, must be determined
  from experiment \cite{(Eks13),(Eks15)} or eventually from LQCD \cite{(Bar15)}. 

A powerful suite of {\em ab-initio}
approaches based on inter-nucleon interactions provides a quantitative description of light and medium-mass nuclei
\cite{(Car15),(Hag14),(Bar13),(Her13),(Lah14),(Som14),(Hag16)} and  their reactions
\cite{(Lov14),(Hup15),(Elh15)}. For
medium-mass  systems, global  configuration-interaction (CI)
methods employing microscopic effective (i.e., medium mediated) interactions \cite{(Bog14),(Jan14),(Sim16),(Shi16),(Tsu16)} offer detailed descriptions of nuclear excitations
and  decays. Modern continuum shell-model approaches unify
nuclear bound states with resonances and scattering continuum within
one consistent framework \cite{(Mic09),(Hag12),(Pap13)}.

For heavy complex nuclei the tool of choice is
the nuclear density functional theory (DFT)
\cite{(Dru10),(Dob11),(Rin11),(Erl12),(Gor13),(Afa13)}; the  validated global  energy density functionals  often provide a level of accuracy
typical of phenomenological approaches based on parameters locally
fitted to the data, and enable wide extrapolations into nuclear {\it terra incognita}.
The time-independent and time dependent DFT extensions have provided quantitative description of one of the toughest problems of nuclear structure: nuclear large amplitude collective  motion, which includes such phenomena as shape coexistence, fission, and heavy-ion fusion 
 \cite{(Sim14),(Uma15),(Bul15),(Sad16),(Sat16),(Mat16)}.

Complex nuclei often display regular patterns indicating the presence of emergent phenomena, such as collective rotation, superfluidity, and phase transitions. At the energy scale of collective excitations of the nucleus, collective models based on geometric concepts or dynamical symmetries associated with specific algebras are extremely useful, for they are capable of describing vasts amounts of experimental data using very few parameters \cite{(Cas06),(Coe15),(Isa16)}. 
The  perspective offered by such ``low-resolution" models is complementary to the microscopic view of a nucleus made from small building blocks.  

The important challenge for the field is to connect different many-body approaches, describing the nucleus  at different resolution scales,  in the regions of the nuclear landscape where they overlap. By bridging the gaps, one is aiming at developing one comprehensive  picture of the atomic nucleus, from the nucleon all the way to superheavy elements \cite{riatheory,(Ber07)}. 
Such multiscale approaches can be found in all areas of science. 
The emphasis on  {\it resolution scale} is important: the resolving power of a theoretical model  should always be as low as reasonably possible for the question at hand \cite{(Fur10),(Fur14)}. Indeed, there is no point  in explicitly involving quarks and gluons when dealing with the low-lying excitations of $^{120}$Sn, and  there is no need to worry about the short-range behavior of nuclear force when studying the low-energy transfer. Or, as formulated by Weinberg in his
Third  Law of Progress in Theoretical Physics:
``You may use any degrees of freedom you like to describe
a physical system, but if you use the wrong ones, you will be sorry!"
\cite{(Wei83)}. 

\begin{figure}[htb]
\center
  \includegraphics[width=0.7\textwidth]{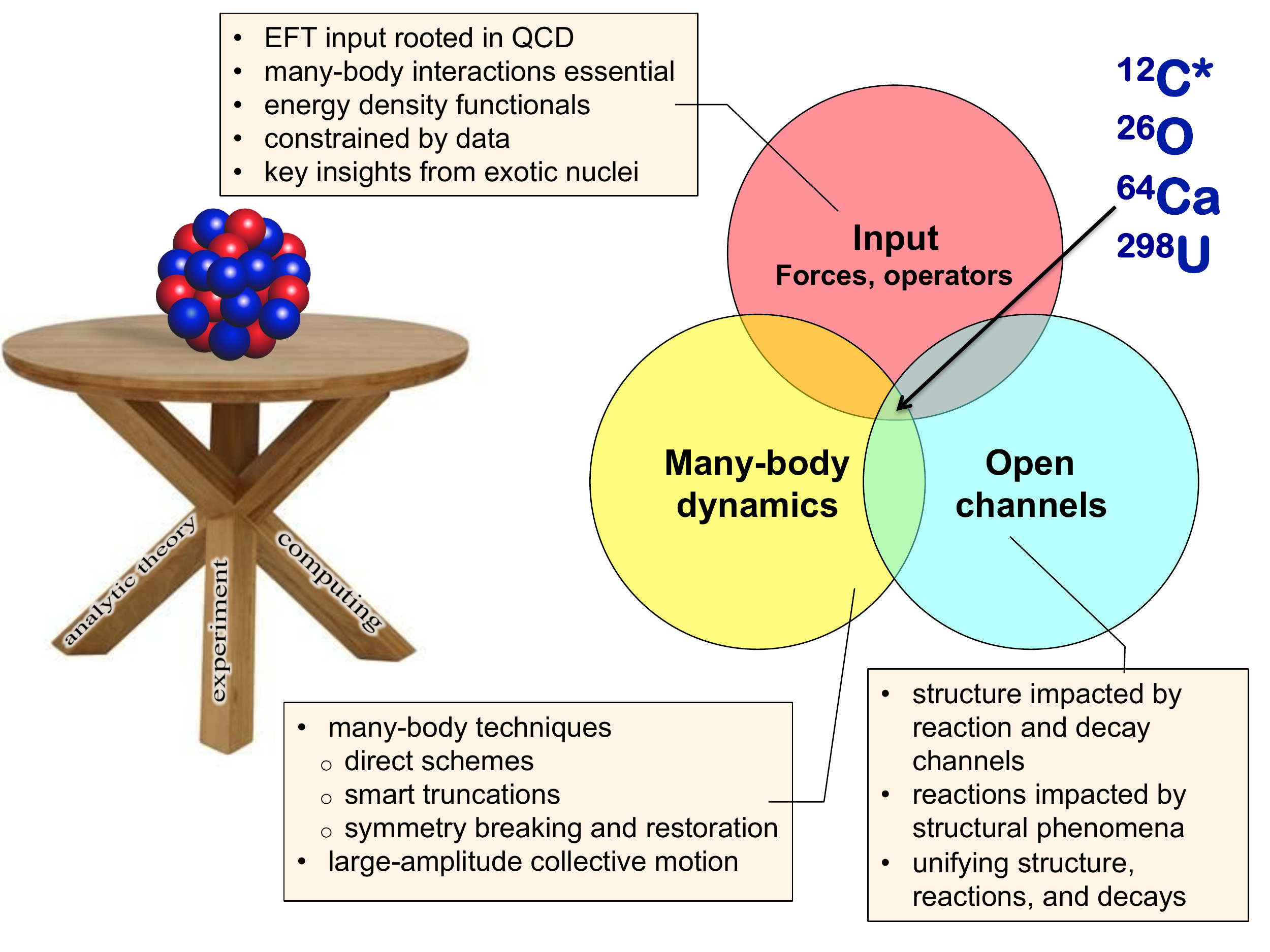}
\caption{\label{challenges} Left: Three pillars of modern research with nuclei: experimentation, analytic theory, and computer simulations. Right: Challenges of the nuclear many-body problem.
A comprehensive theoretical framework that
would be quantitative, have predictive power, and provide uncertainty quantification 
must meet three stringent requirements: (i) the input (interactions and operators) must be quantified and of high quality; (ii) many-body dynamics and
correlations must be accounted for; and (iii) the associated formalism
must take care of open-quantum-system aspects of the nucleus.
Only then can one hope to understand important nuclear states, such as the Hoyle state in $^{12}$C, two-neutron emitter $^{26}$O, weakly bound or unbound 
$^{64}$Ca, and $^{298}$U (r-process fission cycling system). 
}
\end{figure}
The hierarchy of nuclear models describing  the nucleus at different resolving power, shown in Fig.~\ref{DoF},  has one common feature: all those approches are phenomenological at some level;  we do not have at our disposal a
well settled ``exact" starting point. That is, in order to describe physical reality, the low-energy coupling constants of nuclear structure models must be optimized to hand-picked experimental data.  
In this context, the superlatives such as ``fully microscopic" or ``from first principles",  used to characterize particular approaches, can hardly  be viewed as absolute. Nuclear theory requires phenomology although we seek to minimize it where possible.

On the other hand, precisely because of the phenomenological nature of nuclear theory, there are tremendous opportunities for constraining and developing  nuclear models provided by
rare isotopes \cite{Transactions,RISAC,(Dob07),(Bal14)}. The next-generation of rare isotope beam  facilities will enable  access to key regions
of the nuclear chart, where the measured nuclear properties will challenge current theory, highlight shortcomings,  and  identify modifications needed. 
The challenge is to develop methodologies to reliably calculate and understand the
origins of unknown properties of new physical systems; physical
systems with the same ingredients as familiar ones but with
new and different properties due to large neutron-to-proton asymmetries 
and low-energy reaction thresholds.  A related
challenge is to be able to identify the impact of new observables,
quantify correlations between predicted observables, and assess
uncertainties of theoretical predictions, see Sec.~\ref{errors}.

New theoretical ideas, key data on rare isotopes,
high fidelity simulations using leadership computing platforms, and
increased interdisciplinary collaborations  \cite{(Bog13),(Bal14)} -- each pave the way for
continued  progress in theoretical nuclear structure studies. Figure~\ref{challenges} shows, symbolically, three pillars of modern research with nuclei: experimentation, analytic theory, and supercomputing \cite{Decadal2012}. This figure also illustrates the main challenges faced by the modern nuclear many-body problem:
the need for a validated, high quality input, the 
importance of  many-body dynamics, and the impact of open reaction and decay channels. For  excellent collections of papers on burning questions in nuclear structure and reaction theory, I refer the reader to the two recent Focus Issues of J. Phys. G \cite{(Dob10),(Joh14)}.

\section{How to create and maintain unfair advantage?} 

What can modern nuclear structure theory do -- of course, in addition to continuously creating an influx of fresh  ideas -- to improve its efficiency, to be able to  respond nimbly to opportunities for scientific discovery? One mechanism to create an unfair advantage \footnote{``Always work on problems for which you possess an unfair advantage" (H. Bethe).} is to team up into 
large multi-institutional efforts involving strong coupling between nuclear physics theory, computer science, and applied math. While experimentalists are no strangers to large collaborative efforts (a paper  on the mass of the Higgs boson \cite{Higgs} with 5154 authors has recently broken the record for the largest number of contributors to a research article), nuclear theorists have eventually come to the realization that ``the whole is greater than the sum of its parts."

\begin{figure}[htb]
\center
  \includegraphics[width=0.7\textwidth]{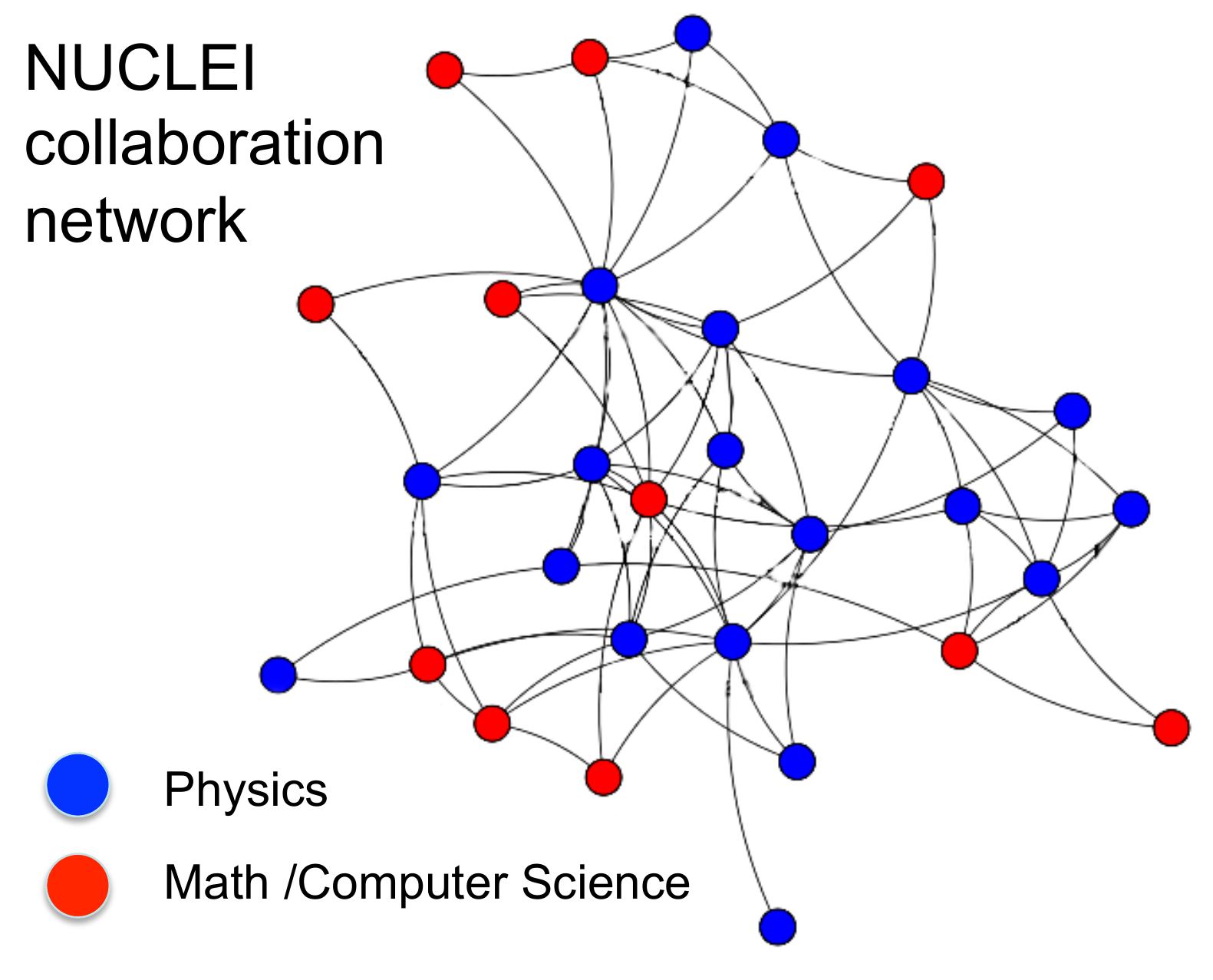}
\caption{\label{links}
Collaboration links within NUCLEI \cite{NUCLEI} illustrating the integration of the project as of 2016. Physics and math/CS principal investigators are marked by blue and red dots, respectively. See Ref.~\cite{NUCLEI} for details. (Based on the plot by Rusty Lusk.)
}
\end{figure}
In this context, one representative example is the NUCLEI collaboration involving nuclear structure  theorists, computer scientists (CS) and mathematicians from 16 institutions, including 10 universities and 6 national labs \cite{NUCLEI}.  The scope of nuclear science and math/CS is quite wide-ranging. Over the course of NUCLEI, and its UNEDF predecessor \cite{(Bog13)}, collaborations across domains  have grown, and now involve many direct connections. Figure~\ref{links} illustrates the present status of collaborations within  NUCLEI.
As is apparent from this network diagram, the math/CS participants are directly embedded in the various nuclear theory  efforts. In each partnership, math/CS participants collaborate with physicists to remove barriers to progress on the computational/algorithmic physics side. This has proven to be a very successful organizational strategy that has resulted in many excellent outcomes, some initially unanticipated. The UNEDF/NUCLEI case emphasizes the importance of assembling
agile theory teams working on important questions and  programmatic deliverables that would be difficult, or impossible, to tackle by individual investigators or atomized small groups. Even more
 opportunities are created by taking nuclear theory research global, see Ref.~\cite{NSACLRP2015} for  examples of international collaborations. 

\begin{figure}[htb]
\center
  \includegraphics[width=0.6\textwidth]{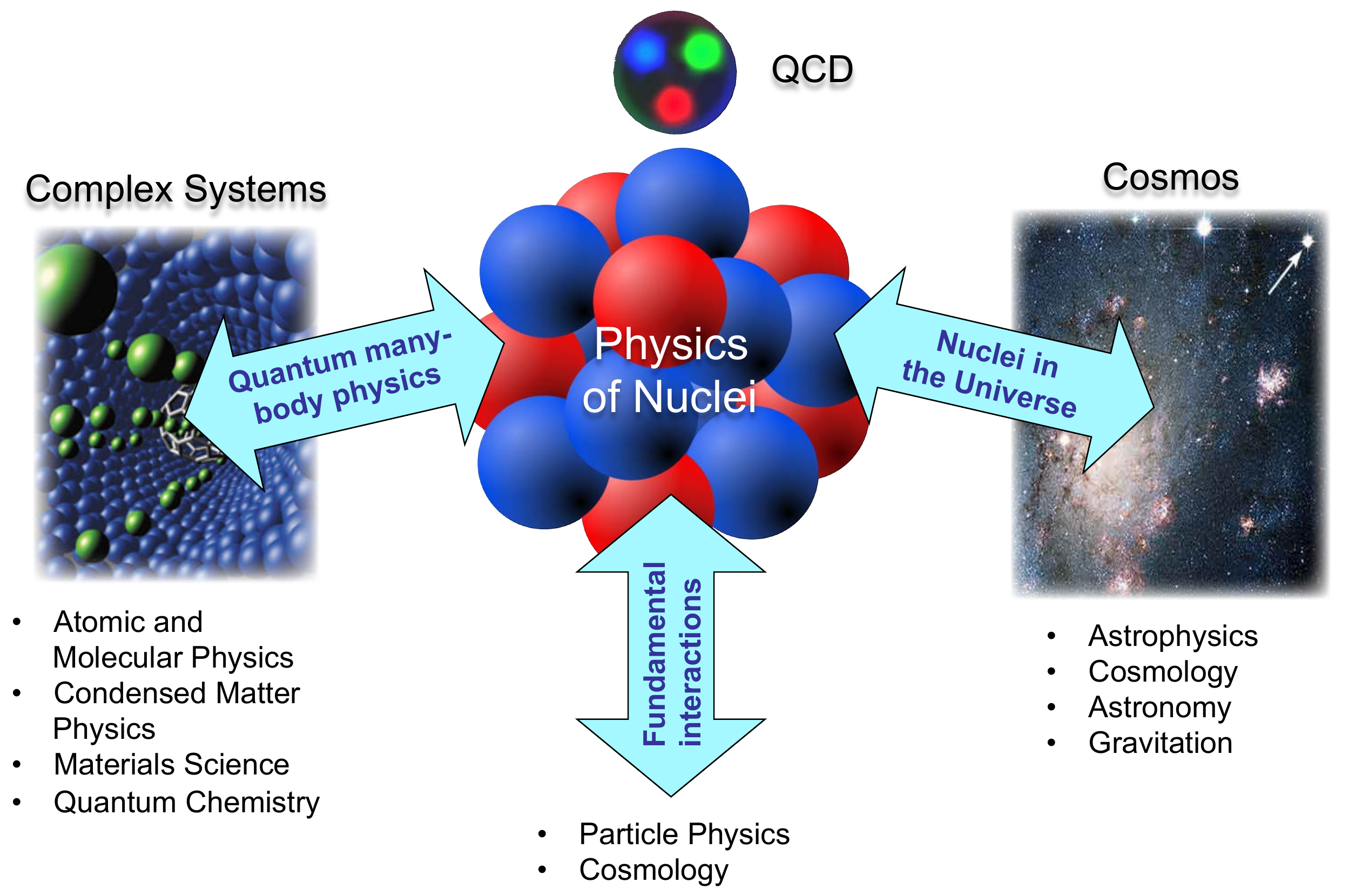}
\caption{\label{intersections}
There are diverse intersections of nuclear theory research with other areas of physics. Examples of research on interfaces include: strongly coupled superfluid systems; phase-transitional behavior; spectral fluctuations and statistics; properties of open quantum systems; clustering; studies of neutron-rich matter as in neutron stars and supernova;  and
nuclear matrix elements for fundamental symmetry tests in nuclei and for neutrino physics. 
(Based on Ref.~\cite{(Bal14)}.)
}
\end{figure}
As illustrated in Fig.~\ref{intersections}, the nuclear many-body problem is  an interdisciplinary enterprise, which connects to many areas of physics  \cite{Decadal2012,NSACLRP2015,NSACLRP2007}. By exploring interfaces across domains, we create opportunities to answer fundamental  questions pertaining to  our field. Historically, nuclear theory has benefited tremendously from such a strategy. For instance, many aspects of the collective model of the nucleus \cite{(Boh69),(Boh75)} follow  general schemes developed in molecular physics. The EFT  and renormalization group techniques born in high energy physics have revolutionized low-energy nuclear structure research \cite{(Bad02),(Fur12)}. Examples abound in all areas of the nuclear many-body problem, see, e.g., Ref.~\cite{(RS80)}. Great facilitators of interdisciplinary interactions are  theoretical nuclear theory centres, such as the ECT* in Trento or the Institute for Nuclear Theory in Seattle. One of the main goals of the recently established  FRIB Theory Alliance \cite{FRIBTA} will be to  reach
beyond the traditional fields of nuclear structure and reactions, and nuclear astrophysics,
to explore exciting interdisciplinary boundaries 
\cite{(Bal14)}.

Many great opportunities in theoretical nuclear structure are, and will be, provided by high-performance computing; the best is yet to come with the anticipated arrival of  exascale platforms. Large-scale simulations  are essential for providing predictive capability, estimating uncertainties,  assessing model-based extrapolations, and realizing the scientific potential of current and future experiments, see Fig.~\ref{challenges}. Here, significant work is needed to take advantage of
extreme-scale capabilities as they become available, in particular to develop the needed multi-disciplinary workforce in computational nuclear theory. As illustrated in Ref.~\cite{(Hag16)},  even with the raw computational power growing almost exponentially, the overall improvement in nuclear structure modelling  has outstripped this growth due to  improved algorithms and  efficient codes. However, as computational environments become more diverse in architectural design, and more specialized in algorithmic implementation, it is crucial to grow the community of nuclear theorists with skills sufficient to master the challenges provided by extreme-scale computing, in particular the emerging trends in computer architectures.

\section{Quality control}\label{errors} 

\begin{figure}[htb]
\center
  \includegraphics[width=0.8\textwidth]{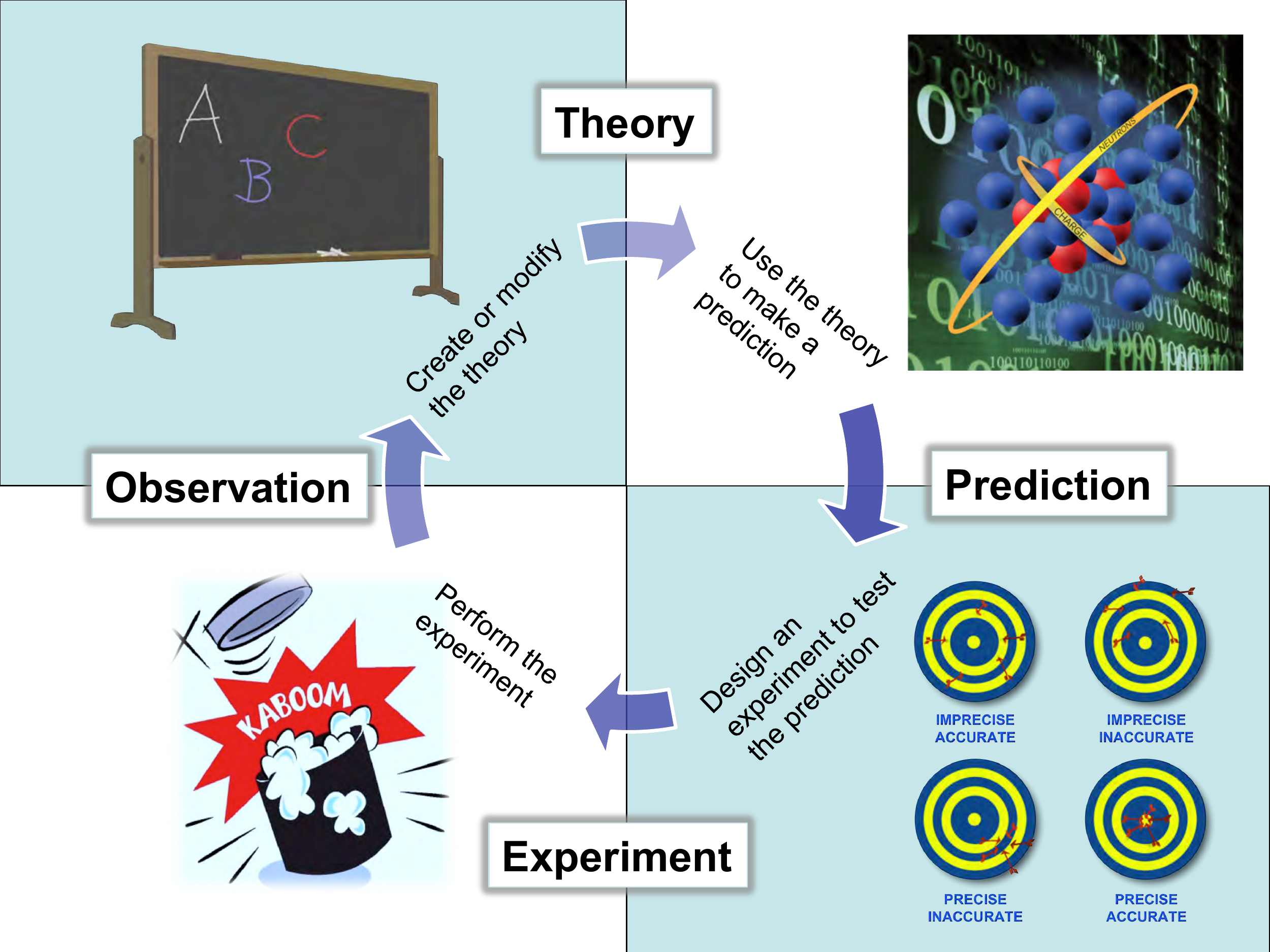}
\caption{\label{Scientificmethod} 
Schematic illustration of the scientific method as applied to nuclear structure research. (Based on Ref.~\cite{(Ire15)}.) Exact models are seldom available in realistic nuclear structure theory. At the heart of modern developments  is the uncertainty quantification of theoretical predictions. The basic question is: what is the best way to use experimental data in the formulation of theoretical models that attempt to explain the results of experiments and make predictions for new observables, often involving huge extrapolations? 
}
\end{figure}
As in other  areas of science, nuclear structure  uses a 
cycle of ``observation-theory-prediction-experiment-", to investigate phenomena, build  knowledge, and define future research.
Such an approach, known
as the scientific method,  guides the relationship between theory
and experiment: theory is modified or rejected  based on new experimental data and the improved theory can be used to make predictions that guide future measurements.

The positive feedback in the experiment-theory cycle illustrated in a schematic way in Fig.~\ref{Scientificmethod}  can be enhanced if statistical methods and computational methodologies are applied
to determine the uncertainties  of model parameters and calculated observables. In partnership  with applied mathematics and computer science,
modern nuclear structure theory strives to estimate errors on predictions
and assess extrapolations. This is essential for  developing predictive
capability, as theoretical
models are often applied to entirely new nuclear systems and conditions that are not accessible to experiment. Statistical tools can be used to both improve and eliminate a model, or better define the range of a model's validity. In this context, it is important to realize that: (i) it makes no sense to improve a model below that model's resolving level;  (ii) a model can be very well determined and yet very wrong; and (iii) model's error budget contains information on what data  are needed to make further improvements.
For some relevant recent work, see 
Refs.~\cite{(Rei10),(Fat11),(Gao13),(Kor13skins),(Dob14),(McD15),(Roc15),(Car15a),(Fur15)}.

Because of the phenomenological nature of the field, quality input is essential. It makes little sense to waste precious human time and computer cycles on sophisticated simulations, if the input (interactions, operators) has not been validated. Or, in other words, we want to avoid a ``garbage in, garbage out" situation. The need for uncertainty estimates of theoretical models has been recognized in the nuclear physics community. So the question is not whether to do it or not, but how to do it best \cite{(ISNET)}.

\section{Looking into the crystal ball}

As once remarked by Niels Bohr, in the context of  the Heisenberg Uncertainty Principle,  ``It is exceedingly difficult to make predictions, particularly about the future." But, judging by the rate of current progress in theory of nuclei, we are  well positioned  to anticipate long-term developments. (It is interesting to note that many requirements for nuclear structure theory postulated in  2005 \cite{(Bar05)} have already been met.)
So here follows the list of anticipated,  more-than-a-gut-feeling, Ten Nuclear Structure Theory  Greatest Hits for the next 10-15 years.
\begin{enumerate}
\item
We will describe the lightest nuclei  in terms of lattice QCD and, by the way of coupling with EFT, we will understand the QCD origin of nuclear forces. 
\item
We will develop a  predictive framework for light, medium-mass nuclei, and nuclear matter from 0.1 to twice the saturation density, rooted in realistic inter-nucleon interactions.  {\it Ab-initio}  methods will reach  heavy nuclei in the next decade.
\item
We will develop predictive and quantified nuclear energy density functional rooted in {\it ab-initio} theory. This spectroscopic-quality  functional will properly extrapolate in mass, isospin, and angular momentum to provide predictions in the regions where data are not available.
\item
We will provide the microscopic underpinning of  collective models that explain dynamical symmetries and simple patterns seen in nuclei. In this way, we will link fundamental and emergent aspects of nuclear structure.
\item
By developing  many-body approaches to  light-ion reactions and large-amplitude collective motion, we will have at our disposal predictive models of fusion and fission that will provide the missing data for astrophysics, nuclear security, and energy research.  
\item
By exploring quantum many-body approaches to open systems, we  will understand the mechanism of clustering and explain properties of key cluster states and cluster decays.
\item
By taking advantage of realistic many-body theory, we will unify the fields of nuclear structure and reactions. 
\item
We will achieve a comprehensive description, based on realistic structural input,  of nuclear  reactions with  complex projectiles and targets, involving  direct, semi-direct, pre-equilibrium, and compound processes.
\item
We will carry out predictive and quantified calculations of nuclear matrix elements for fundamental symmetry tests in nuclei and for neutrino physics. This will require exploring the role of correlations and currents.
\item
By taking the full advantage of extreme-scale computers, we will master the tools of uncertainty quantification. This will be essential for enhancing the coupling between theory and experiment -- to provide predictions that can be trusted.
\end{enumerate}
It will be instructive to revisit this list in ten years, on the 50th anniversary of the Journal!

\section{Concluding remarks} 

\begin{figure}[htb]
\center
  \includegraphics[width=0.7\textwidth]{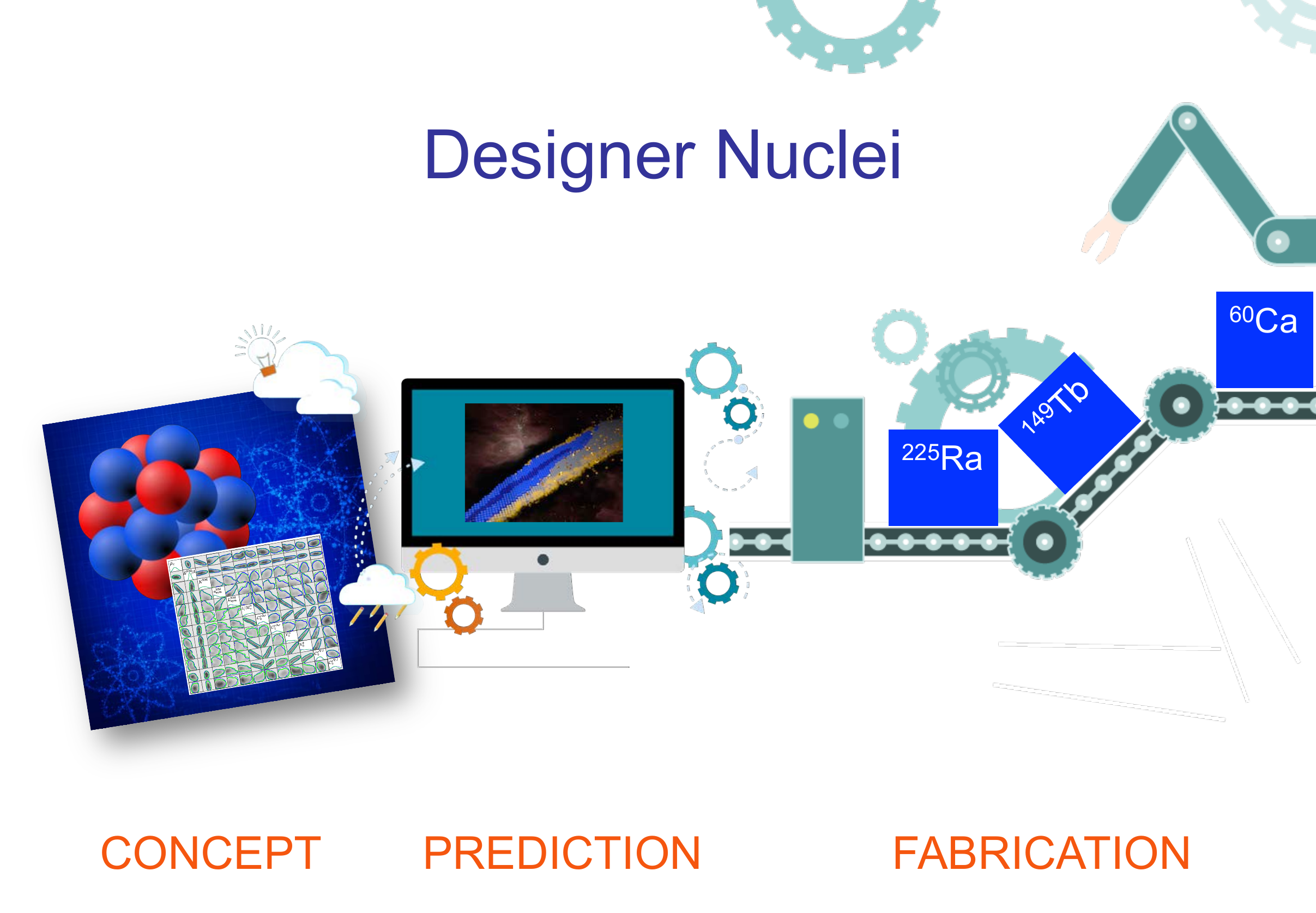}
\caption{\label{designer} 
Some nuclei are more useful than others, either for basic science or for applications.  Nuclear theorists and experimentalists  are getting increasingly better in drafting, predicting, fabricating, and characterizing  femtostructures with desired properties, the designer nuclei \cite{(She08),(Jon10)}. In this process, theory and experiment go hand in hand. Three such designer nuclei are illustrated: $^{225}$Ra, which can potentially  explain the dominance of matter over antimatter in the present Universe \cite{(Par15)}; $^{149}$Tb, an ``alpha-knife" radionuclide that kills cancer cells \cite{(Bey04)}; and $^{60}$Ca, which is crucial for understanding the limits of nuclear existence \cite{(For13)}. 
}
\end{figure}
We are witnessing a renaissance of nuclear structure research due to  a revolution in experimentation, analytic theory, and computing. The result is a shift from a phenomenological picture of the nucleus to nuclear theory grounded in the Standard Model. During the last decades we sketched  a  roadmap for a predictive theory of nuclei. Today, the journey has begun.

On this journey,  important milestones will be marked by designer nuclei with characteristics adjusted to specific research needs \cite{(She08),(Jon10)}, see Fig.~\ref{designer}.
The goal for theory  is  to be able to determine which nuclei are  most important and  describe  their properties quantitatively. Some designer nuclei will be fabricated and characterized in experimental  laboratories; these will provide crucial anchor points for nuclear models.  Some will be predicted by theory and characterized experimentally. Some designer nuclei, 
inhabiting  remote regions of the nuclear landscape that are not accessible by experiment, will be only accessed by theory. 
With a picture of nuclei based on the correct microphysics and the data from rare isotopes, we will develop the comprehensive model of the atomic nucleus: 
to understand, predict, and use.

\bigskip   
I would like to thank many colleagues, in particular    Jacek Dobaczewski, Rick Casten, David Dean, Dick Furnstahl, Morten Hjorth-Jensen, Thomas Papenbrock, Marek P{\l}oszajczak,  and Paul-Gerhard Reinhard for many   stimulating discussions. Graphic arts assistance of Erin O'Donnell is gratefully acknowledged.
This work was supported by the U.S. Department
of Energy, Office of Science, Office of Nuclear Physics under Award
Numbers DE-SC0013365 (with Michigan State University) and   DE-SC0008511 (NUCLEI SciDAC-3 collaboration).

\section*{References}

\end{document}